\documentclass[a4paper,twoside]{article}

\usepackage{epsfig}
\usepackage{subcaption}
\usepackage{graphbox}
\usepackage[hidelinks]{hyperref}
\usepackage{cite}
\usepackage{calc}
\usepackage{amssymb}
\usepackage{amstext}
\usepackage{amsmath}
\usepackage{amsthm}
\usepackage{multicol}
\usepackage{pslatex}
\usepackage{apalike}
\usepackage{algorithm2e}
\usepackage[bottom]{footmisc}
\usepackage{SCITEPRESS}     % Please add other packages that you may need BEFORE the SCITEPRESS.sty package.

\newcommand{\acro}[1]{\textsc{\MakeLowercase{#1}}}

\begin{document}

\title{The Dance of Logic and Unpredictability: Examining the Predictability of User Behavior on Visual Analytics Tasks}

\author{\authorname{Alvitta Ottley\orcidAuthor{0000-0002-9485-276X}}
\affiliation{Washington University in St.\ Louis, USA}
\email{alvitta@wustl.edu}
}

\keywords{intelligent visual analytics; artificial intelligence; human-machine collaboration; individual difference; \\user modeling;}

\abstract{The quest to develop intelligent visual analytics (\acro{va}) systems capable of collaborating and naturally interacting with humans presents a multifaceted and intriguing challenge. \acro{va} systems designed for collaboration must adeptly navigate a complex landscape filled with the subtleties and unpredictabilities that characterize human behavior. However, it is noteworthy that scenarios exist where human behavior manifests predictably. These scenarios typically involve routine actions or present a limited range of choices.
This paper delves into the predictability of user behavior in the context of visual analytics tasks. It offers an evidence-based discussion on the circumstances under which predicting user behavior is feasible and those where it proves challenging.  
We conclude with a forward-looking discussion of the future work necessary to cultivate more synergistic and efficient partnerships between humans and the \acro{va} system. This exploration is not just about understanding our current capabilities and limitations in mirroring human behavior but also about envisioning and paving the way for a future where human-machine interaction is more intuitive and productive.
%at least 70 and at most 200 words. The text must be set to 9-point font size.
}

\onecolumn \maketitle \normalsize \setcounter{footnote}{0} \vfill

\section{\uppercase{Introduction}}
\label{sec:introduction}

Building intelligent visual analytics systems that can assist and interact with humans during data analysis is akin to teaching a robot to dance. We aspire to achieve a dance of data with a fluid exchange of ideas, a graceful understanding of needs, and a seamless partnership in pursuing hypotheses, insights, and decisions. However, the human element in this equation is far from a predictable automaton – humans are complex, driven by emotions, experiences, and social contexts that often elude the straightforward logic of machines. This complexity presents the visual analytics community with a formidable challenge: \textit{How do we design systems that intelligently collaborate with their human counterparts?}

As we examine examples beyond visual analytics, a common misconception frames humans as purely logical entities whose decisions and actions are easily predictable by well-defined rules. 
This assumption is evident in technologies like basic customer service chatbots, which are programmed for simple inquiries~\cite{sheehan2020customer} but falter with complex or emotionally charged interactions~\cite{prentice2020engaging}, resulting in unhelpful customer experiences~\cite{chong2021ai,crolic2022blame,huang2022chatbots}. Similarly, advertising algorithms that target based on demographics and past behaviors often fall short~\cite{white2019programmatic}. 
They assume that human preferences are static, overlooking the subtleties of an individual's goals and ever-evolving needs~\cite{lambrecht2013does}. Consequently, these approaches can lead to irrelevant, intrusive, or untrustworthy advertising~\cite{bleier2015importance}. 
Moreover, the frequent shortcomings of these systems can largely be attributed to their inability to cope with the broad spectrum of unpredictable factors inherent in the given case scenarios. Adopting a one-size-fits-all strategy fails to consider the unique variances among individuals and the significant role that emotional factors play in shaping human decisions and preferences under those circumstances~\cite{bleier2015importance}.

\begin{figure*}[t]
    \centering
     \begin{subfigure}[c]{0.5\textwidth}
         \centering
         \includegraphics[width=\textwidth]{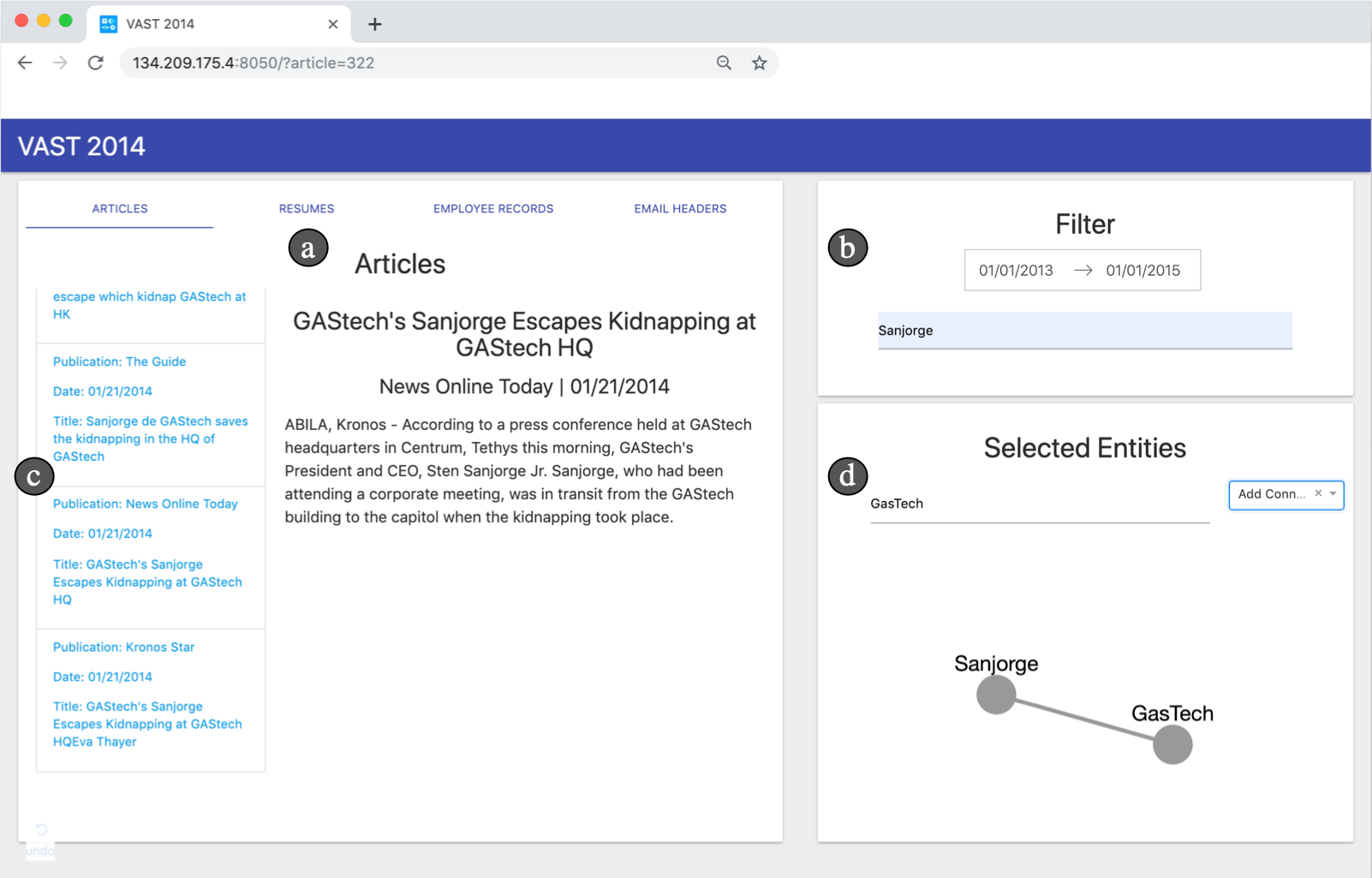}
         \caption{{\small The interface used to solve a task related to a kidnapping crime. Users can (a) view details, (b) filter, (c) list matching results, and (d) sketch an entity/connection network.}}
         %\label{fig:y equals x}
     \end{subfigure}
     %\hfill
    \raisebox{0.003\height}{
     \begin{subfigure}[c]{0.4\textwidth}
         \centering
         \includegraphics[align=c, width=\textwidth]{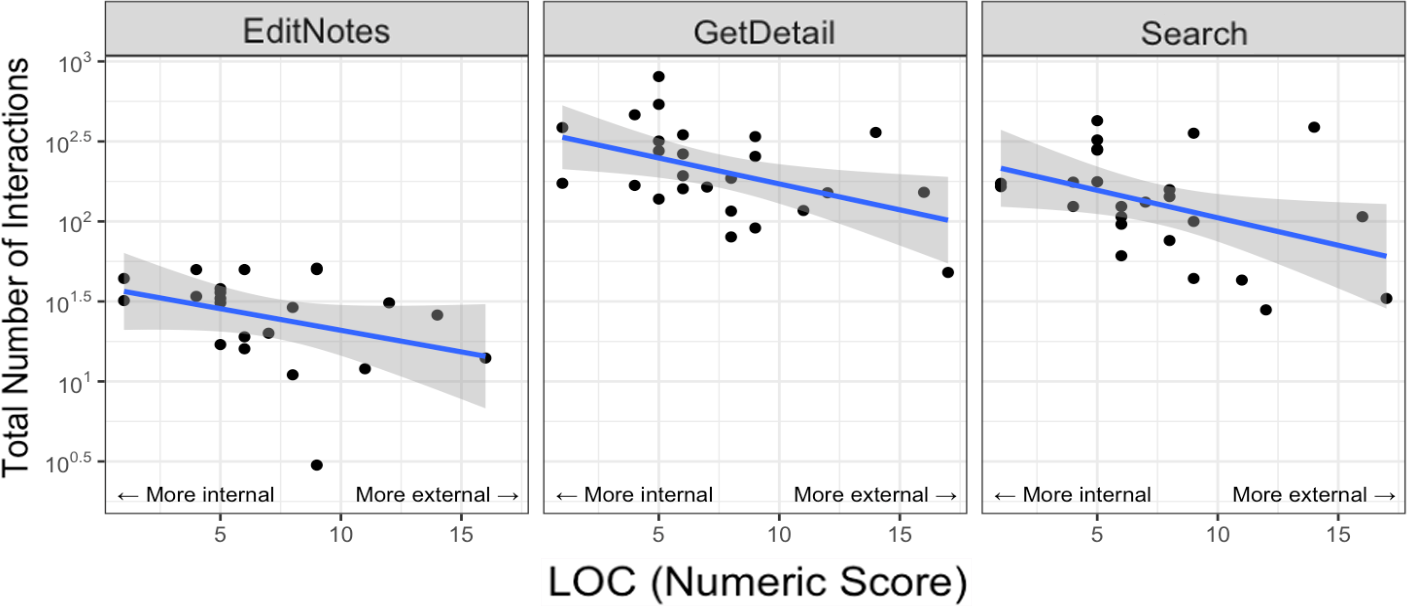}
         \caption{{\small The analysis revealed correlations between the number of interactions and participants' locus of control across the three primary action types.}}
    \end{subfigure}
    }
    \caption{The interface and analysis result from Crouser et al. They analyze the analysis behaviors from a series of exercises with 22 trained intelligence analysts~\cite{crouser2020investigating}. Their preliminary analysis suggests that individual differences in locus of control can modulate expert behavior in complex analysis tasks.}
    \label{fig:las_study}
\end{figure*}

Yet, in certain situations, human behavior tends to be predictable~\cite{heiner1983origin,flanagan2003action}. These situations usually involve routines, repetitive actions, or limited choices. 
For example, many people have regular commuting patterns. Most people's daily routines have only slight variations~\cite{krumme2013predictability}, making this predictability useful for traffic forecasting and scheduling public transportation~\cite{song2010limits}.
People's interaction with basic technology, like ATMs or elevators, tends to follow a formulaic script due to the limited actions available. Further, purchasing patterns for essential goods often show consistency~\cite{kim1997studying,krumme2013predictability}. %These scenarios highlight instances where human behavior demonstrates a high level of predictability. %However, the challenge lies in accurately identifying these scenarios and designing systems that respond appropriately.

This paper argues that visual analytics systems can capitalize on the predictable aspects of human behavior. 
This could mean creating interfaces and functionalities that cater to routine tasks while providing the flexibility and depth required for more complex, less foreseeable analytical endeavors. For instance, if a system recognizes that users frequently perform a specific sequence of actions, it can automate or simplify these steps. This approach could enhance efficiency and minimize the user's cognitive load, allowing them to focus on more complex data analysis aspects requiring deeper thought and creativity.

However, the challenge lies in discerning when human actions are routine and predictable and when they are not. 
This balance is key to developing visual analytics systems that are truly collaborative partners in the dance of data exploration and analysis.
This paper discusses some necessary steps for creating intelligent visual analytics tools:
\begin{itemize}
    \item A deeper collaboration between humans and \acro{AI} requires embracing the complexity of human behavior. We discuss the role of individual differences in visual analysis in Section~\ref{sec:individual_differences}.

    \item Section~\ref{sec:va_impact} explains how the system's design can affect action predictability. 

    \item Two case studies in Section~\ref{sec:casestudies} demonstrate action prediction based on user interactions.

    \item In Section~\ref{sec:framework} advocates for broadening the conceptual models of human-machine collaboration in visual analytics. We suggest a framework that integrates \acro{AI} capabilities with human expertise.

    \item Finally, we discuss, among other things, the ethical considerations for human-AI interactions that must be rigorously addressed. 
    
\end{itemize}

\section{The Interplay of Predictability and Individuality in Data Analysis}
\label{sec:individual_differences}

While it is true that certain scenarios can lead to predictable behavior patterns, this does not negate the rich tapestry of individual differences that manifest in various ways during visual analysis tasks. These differences are influenced by many factors, including personality traits, cognitive abilities, and situational conditions, each playing a significant role in how individuals interact with and interpret data~\cite{ottley2022adaptive,liu2020survey}.

Studies show several individual differences have consistently impacted performance, as evidenced by multiple independent researchers' replication in various experimental settings~\cite{ottley2022adaptive,liu2020survey}.
Personality traits, for example, can greatly influence how a user approaches a visual analytics task. A notable instance is the influence of locus of control, which reflects an individual's perception of control over external events and often affects the speed and accuracy of visualization tasks~\cite{ottley2015manipulating,ziemkiewicz2012visualization}. This impact has been observed across studies using diverse datasets and methodologies and in traditional laboratory experiments and crowdsourcing research platforms~\cite{crouser2020investigating,ottley2015personality}.

For a concrete example, we examine one particular study, described in \autoref{fig:las_study}, examining the behavior of 22 Navy Reservists during complex analytical tasks revealed a correlation between locus of control and expert behavior~\cite{crouser2020investigating}. It found that participants with a more internal locus of control engaged in more actions and covered more data in the same timeframe. Additionally, other studies underscore the importance of visualization design in this dynamic, showing that an individual's locus of control can significantly influence their search strategy in hierarchical systems.

Similarly, cognitive abilities like spatial reasoning, perceptual speed, and working memory capacity can impact the speed and accuracy with which different users understand and analyze complex visual data~\cite{liu2020survey}. Situational factors, including time constraints, task complexity, and the user's emotional state during analysis, further affect this process~\cite{bancilhon2023combining}. Under time pressure, users may adopt heuristic analysis methods, whereas more relaxed conditions might encourage deeper exploration~\cite{bobadilla2018fast,del2016decision}. Moreover, a task's inherent complexity can elicit varying responses, depending on the user's preference for challenge or simplicity~\cite{ziemkiewicz2012visualization}.

After reviewing the research, several key themes emerged regarding the impact of individual differences on visual analytics tasks:

\begin{enumerate}
    \item Individual differences are more pronounced in complex tasks, with greater freedom for exploration~\cite{ziemkiewicz2012visualization,brown2014finding,ottley2015personality}.
    \item Conversely, simpler tasks tend to show less variation in user behavior. Studies often report no substantial effect of individual differences on simpler tasks~\cite{ziemkiewicz2012visualization}.
    \item Even with observable differences between individuals, there are often common behavioral patterns within groups, indicating that for a giving visualization scenario, certain analysis paths are more frequently traversed, even in scenarios with the potential for wide exploration diversity~\cite{brown2014finding,ottley2015personality}.
\end{enumerate}

Understanding individual differences can provide insight into inconsistent and consistent behavior patterns. This knowledge can help create visual analytic tools that intelligently collaborate and respond based on these differences and the situations in which they occur. It respects both the complex nature of human behavior and can improve the functionality of visual analytics systems.

\begin{figure*}[ht]
    \centering
    \includegraphics[width=.8\linewidth]{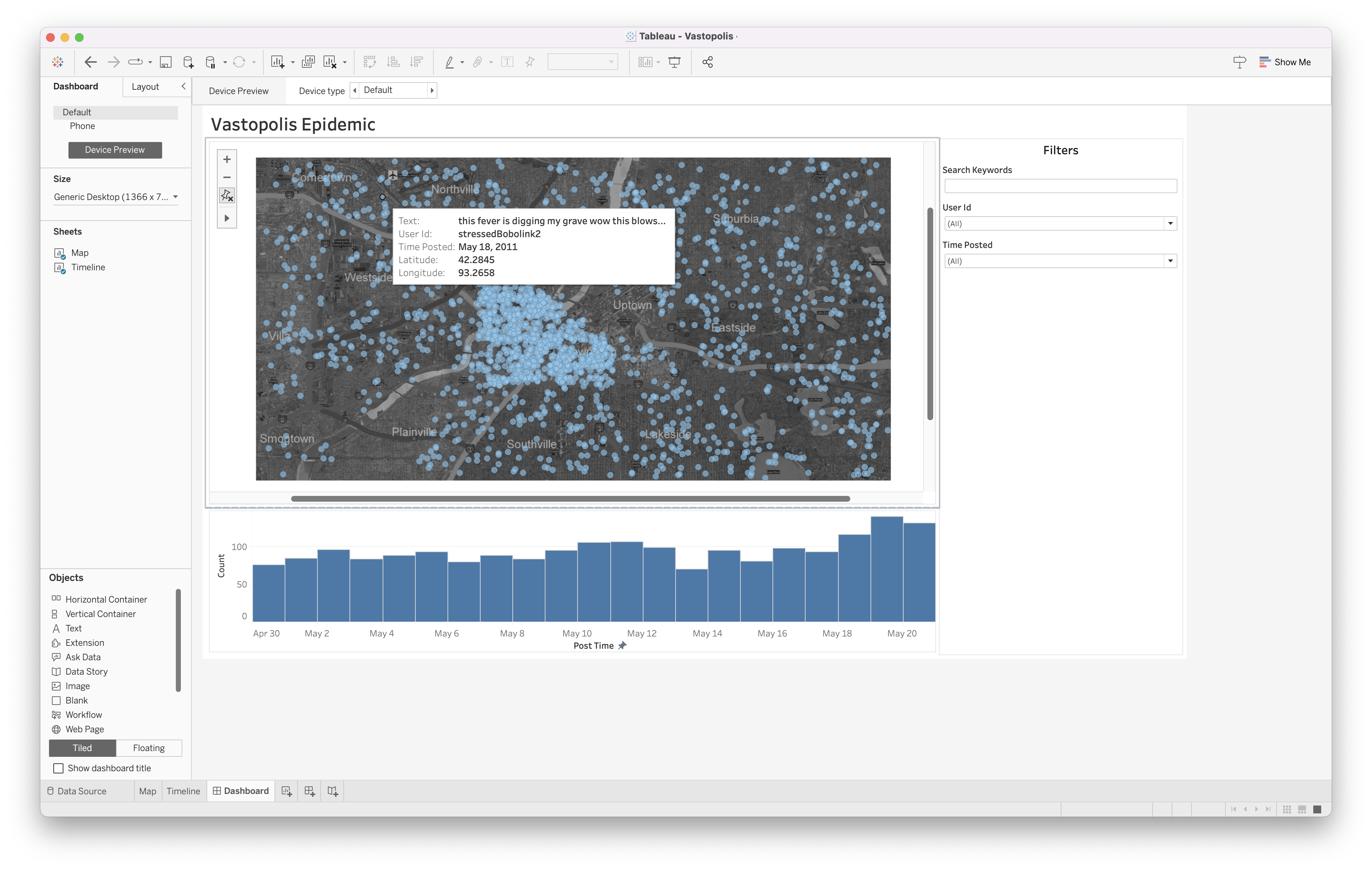}
    \caption{The Tableau interface with a prototype dashboard with an epidemic data set in the fictitious city of Vastopolis, used as the running example in \autoref{sec:va_impact}.  The text displays a map of social media posts with geolocation, a search and filter sidebar, and a bar chart indicating post frequency over three weeks.}
    \label{fig:tableau}
\end{figure*}

\section{How the Design of Visual Analytics Interface Impacts Predictability}
\label{sec:va_impact}

In addition to the analyst's characteristics, the interface design, the nature of the data, and the task at hand can all greatly influence the predictability of user behavior in these \acro{va} scenarios.
Well-designed interfaces typically guide user behavior into predictable patterns by offering clear options and intuitive paths for data exploration, whether intentionally or not. In contrast, a disorganized layout may result in erratic and unpredictable exploration paths, potentially leading users to overlook essential insights and complicating user behavior prediction. 

Consider the dashboard in \autoref{fig:tableau}, which features a simple exploratory interface for analyzing a geospatial dataset. The most dominant feature is a map, occupying roughly seventy percent of the screen. This design choice naturally focuses the user’s attention primarily on the map’s data points. Additionally, given that users typically read from left to right, the filtering options on the right side will likely be the next focus point, followed by the bar chart at the bottom. Thus, predicting attention and high-level areas of interest is feasible.

Interaction affordances, which suggest possible actions through design, also play a crucial role. For example, the persistent visibility of filtering options in \autoref{fig:tableau}, instead of their placement in hidden menus, increases the likelihood of their usage. Users will likely engage with the most accessible actions, such as hover effects, more frequently. Other interactions, like panning and zooming on the map, brushing on the timeline, or clicking on data points, are less obvious due to the absence of explicit visual cues and might be underutilized, especially by new users. The space of possible actions for this interface is small. One might consider using a probabilistic approach to predict action for this interface, encoding the assumed likelihood of observing a specific action as priors and calculating the posterior probability of observing an action given a set of observations.

Additionally, how data is represented dictates the questions an analyst can ask and what they will likely notice and consider. In \autoref{fig:tableau}, the interface's focus on geographical data through maps encourages the exploration of spatial patterns and regional clusters and differences. Similarly, the prominently displayed timeline and area chart at the bottom of the interface are likely to prompt questions about temporal changes. The available filtering options and zooming capabilities influence the depth and specificity of the questions an analyst can pose. An interface that supports intricate data manipulation enables analysts to formulate and test detailed hypotheses, while a more static interface or those without interaction cues might confine them to basic, surface-level observations. Predicting objectives and tasks will require a mapping between them and the observable actions and their association with the current area of interest\cite{gathani2022grammar}.

Now, suppose instead, we consider the interactions more broadly in the Tableau interface or with other advanced statistical analysis tools. This would offer more opportunities to delve into complex questions about correlations or predictions. Moreover, the ability to customize the interface or create custom visualizations significantly broadens the range of potential questions. Analysts are not confined to predefined views and can adapt their analysis to meet specific and unique investigative needs. Furthermore, the amount of data the interface can handle also influences the questions that can be pursued. Some interfaces, optimized for large datasets or real-time data, facilitate queries about broad trends or immediate insights, while others are more suitable for detailed analysis of smaller datasets. Although a more complex system offers greater flexibility, increased degrees of freedom will decrease the predictability of user behavior.

Overall, the interface design implicitly provides guidance or scaffolding to shape the queries, analysis pathways, and questions an analyst considers. This is especially significant for novice users still learning which questions can be asked about data or how to use the system. 
%The interface design shapes how data is perceived and understood and the questions' depth, breadth, and complexity that can be effectively explored and answered. 
Understanding these design elements is crucial for developers of visual analytics systems to create interfaces that facilitate data comprehension and guide users by observing predictable and insightful data interactions.

\section{Case Studies from Visualization Research}
\label{sec:casestudies}

Prior work in visual analytics has demonstrated actions and scenarios where behavior prediction was largely successful and the machine learning techniques used to make these inferences~\cite{xu2020survey}.  This section highlights two such papers. 

\begin{figure}[!ht]
    \centering
    \includegraphics[width=\linewidth]{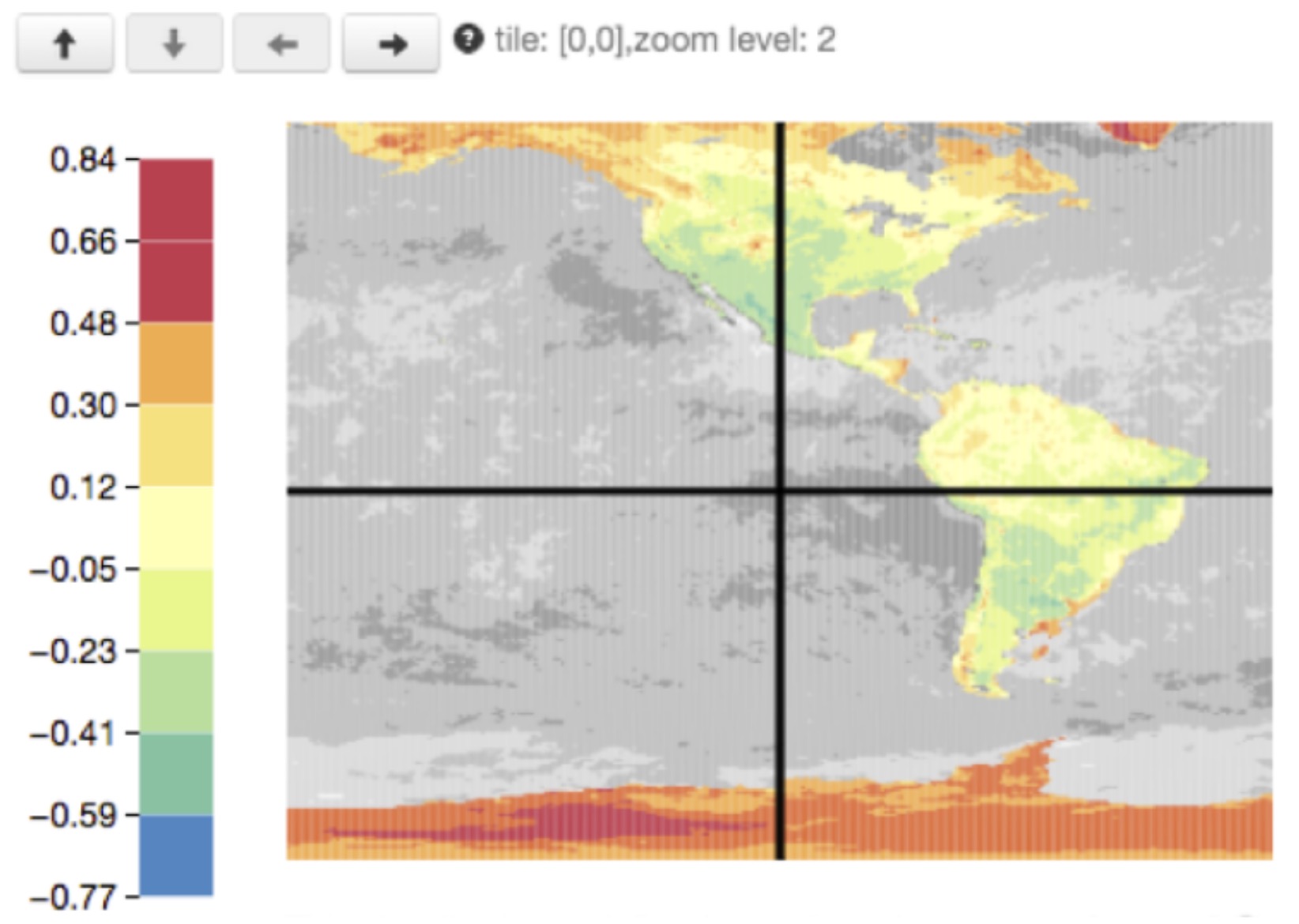}
    \caption{The ForeCache project interface, which visualizes snow levels from NASA MODIS data~\cite{battle2016dynamic}. The authors used observed navigation patterns to predict future interactions and pre-fetch data.}
    \label{fig:forecache-interface}
\end{figure}

\subsection{Predicting Navigation Paths for PreFetching}

Battle et al. explored the feasibility of predicting user navigation behavior to enhance database caching, a valuable feature for managing large datasets with potential latency issues during database queries~\cite{battle2016dynamic}. The project was inspired by previous research highlighted that latency can negatively affect user experience and impede data exploration~\cite{liu2014effects}. To mitigate this, they introduced \textit{dynamic prefetching}, which predicts necessary data to fetch in advance by analyzing users' recent navigation patterns.

\paragraph{Interface}
The research team developed a map-based visualization tool for \acro{NASA}'s \acro{MODIS} snowfall data across America. Given the high-resolution nature of the complete dataset, the system aggregated data into lower-resolution tiles for an overview and increased granularity during user zoom-ins. With its straightforward design, this map interface was conducive to making accurate predictions. It allowed only six observable actions: pan up, down, left, right, and zoom in and out.

\paragraph{Task}
The tasks assigned to the study participants were simple yet effective. Participants were required to explore the data to identify areas with significant snowfall, navigating and searching the interface for regions of interest.

\paragraph{Participants}
The study involved domain scientists, suggesting a uniform background and likely shared expertise. This homogeneity in the participants' backgrounds helped minimize individual differences in skills and knowledge, creating an optimal environment for limiting variability in user interactions.

\paragraph{Predictions}
The researchers used a Markov chain model that predicted users' actions. This model was not pre-programmed but evolved by observing user interactions, enabling the system to learn and update its predictions based on the user’s current state. The evaluation of this dynamic prefetching strategy showed substantial improvements in reducing latency compared to non-prefetching systems (430\% improvement) and significant enhancements in both prediction accuracy (25\% improvement) and latency reduction (88\% improvement) compared to existing prefetching methods.

While the simplicity of this scenario might seem unrepresentative at first glance, it mirrors common situations in data foraging tasks, which are crucial for the sensemaking process~\cite{pirolli2005sensemaking}. Even when multitasking, external actions manifest as sequential rather than concurrent~\cite{mcfarlane1998interruption,mcfarlane2002comparison,mcfarlane2002scope}.
Moreover, divided attention is limited by working memory capacity. Thus, the scope of actions and inquiries at any given time window within a visualization is usually confined. This indicates that the potential for predictive scenarios, like the one in Battle et al.’s study, might be more widespread than initially assumed. Their research provides a solid example of the types of predictions that are feasible – specifically navigation and data foraging – within an accommodating situational environment.

\begin{figure}[b]
    \centering
    \includegraphics[width=\linewidth]{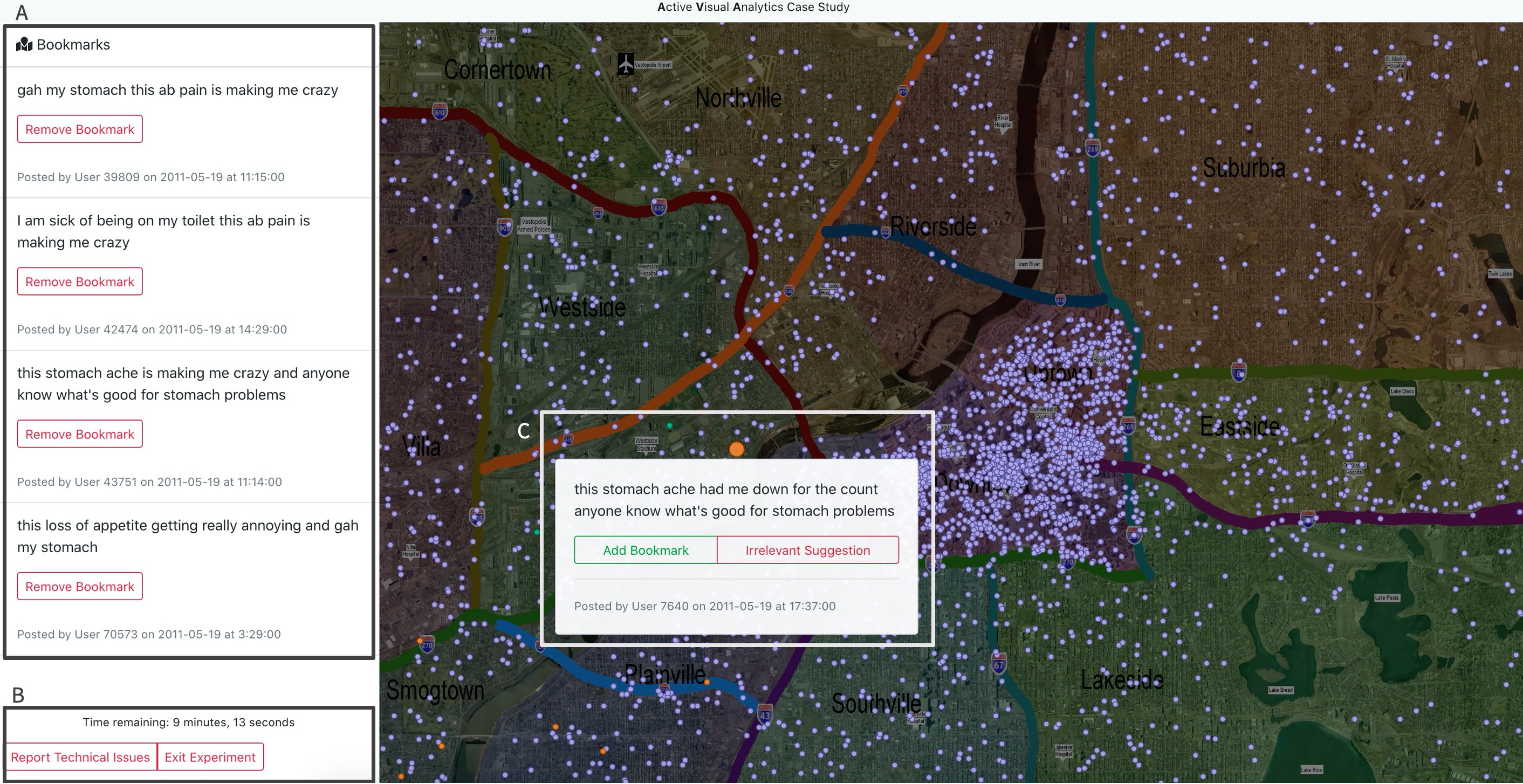}
    \caption{The interface used by Monadjemi et al. in evaluating their algorithm that observes data exploration, infers the relevance of the other points in the dataset and recommends content to the user~\cite{monadjemi2022guided}.}
    \label{fig:enter-label}
\end{figure}

\subsection{Predicting Data Interest for Content Recommendation}

Similarly, Monadjemi et al. aimed to assist data exploration and information foraging. Their approach involved analyzing users' exploration patterns, deducing the characteristics of data points likely to interest the user, and recommending similar points for further exploration. Their primary objective was to expedite data discovery, thereby boosting the efficiency of analytics and enhancing the quality of decision-making.

\paragraph{Interface}
For their evaluation, they adopted a scenario from the VAST 2011 challenge, an annual competition in the visual analytics community focused on addressing real-world challenges. This scenario revolves around a fictional city, Vastopolis, which is grappling with a bio-chemical attack. The authors developed a visualization interface showcasing a city map embedded with geo-tagged social media posts from the past three weeks, providing a comprehensive view of the unfolding situation. Like the ForeCache interface used by Battle et al., the interface was straightforward for limited available actions. The user can pan, zoom the map, and save or unsave relevant social media posts.

\paragraph{Task}
The task assigned to participants was one of reconnaissance and information foraging. Participants were required to explore the data to gauge the range of symptoms being reported on social media. The goal was for them to gather data that downstream analysts could use to understand the extent of the epidemic, assess containment, and hypothesize potential causes. Given the vastness of the dataset, each participant had a ten-minute time limit to identify potentially sick individuals, acknowledging that completing the entire task was beyond expectation.

\paragraph{Participants}
The study involved 130 participants recruited through Amazon's Mechanical Turk platform. These individuals ranged from 18 to 65 years old, were based in the United States, and were proficient in English. While they were not trained analysts, exploring a dataset of social media posts to identify mentions of illness was deemed manageable without specialized training.

\paragraph{Predictions}
The team employed an active search methodology, translating social media posts into numerical values using a standard \textit{word2vec} model and constructing a $k$-NN binary classifier using cosine similarity. As users engaged with the map and bookmarked pertinent posts, the algorithm tagged these data points as relevant. The model continuously updated its understanding of the data after each interaction, reassessing the relevance of unlabeled points in light of recent user actions. It then offered suggestions for additional points the user might explore. 

The analysis of the user study results revealed that the algorithm generated useful recommendations 79\% of the time, on average. Moreover, the data revealed that participants who utilized the algorithm in their search were significantly more efficient than those who did not. The assisted participants discovered a statistically significant greater number of individuals potentially affected by the illness. They also were more adept at distinguishing relevant information from irrelevant data in the dataset.

In summary, Monadjemi et al.'s approach demonstrated the predictability of data interesting in visual analytics, specifically in data foraging tasks~\cite{monadjemi2022guided}. By leveraging machine learning techniques to interpret user interaction and guide further exploration, their system accelerated the data discovery process and enhanced the effectiveness and accuracy of the users' information foraging activities. This study is a testament to the potential of integrating intelligent predictive algorithms into visual analytics systems, paving the way for more intuitive and productive data analysis experiences.

% \subsection{Predicting Attention}

% What do they all have in common, and what do we learn

\begin{figure*}[ht]
    \centering
    \includegraphics[width=\linewidth]{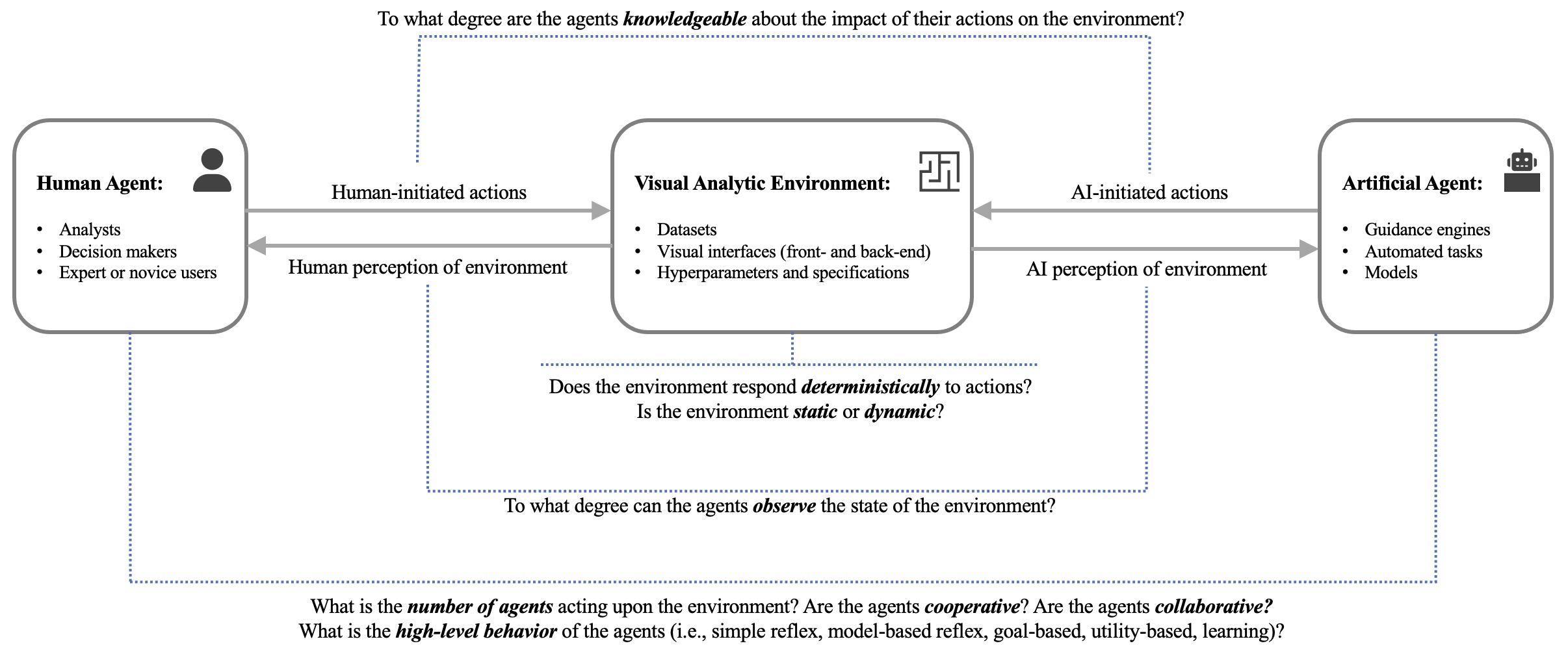}
    \caption{The agent-based framework for visual analytics proposed by \cite{monadjemi2023human}. It adopts terminologies from \acro{AI} and conceptualizes visual analytics scenarios as interactions (observations and actions) between agents and their environment.}
    \label{fig:agent_framework}
\end{figure*}

\section{A conceptual framework for human and AI collaboration}
\label{sec:framework}

The previous section showed successful algorithms that observed human behavior in real-time, predicted actions, and used these inferences to assist the user by recommending exploration or pre-fetching data.
However, to effectively develop collaborative systems, it is crucial to establish a comprehensive framework that recognizes the shared responsibilities and synergistic partnership between human and AI entities~\cite{crouser2013balancing}. Traditional conceptual models in visual analytics have often been limited in scope~\cite{monadjemi2023human}, focusing predominantly on human cognitive processes~\cite{pirolli2005sensemaking}, treating the visual analytic interface as a mere tool without autonomy~\cite{van2005value,van2006views}, or maintaining an imbalanced perspective of the intelligent system, where AI is viewed as having limited capabilities compared to the human's ultimate authority~\cite{sperrle2022lotse,ceneda2017chracterizing}.

However, looking forward to a future where responsibilities are more evenly distributed between humans and AI, broadening these frameworks is imperative. Such an expansion should accommodate the potential for each entity to act as a check and balance against biases that might arise from either side, as proposed by \cite{wall2021left} and \cite{ha2022unified}. Additionally, it's important to consider scenarios involving multiple human and AI agents collaborating on a single task, employing a `divide and conquer' approach.
This revised framework must account for the dynamic interactions between humans and AI, recognizing the unique strengths and limitations of each. In doing so, we can foster systems where collaboration is about task division and mutual learning and support, leading to more robust and effective problem-solving strategies.

%By acknowledging and integrating these considerations, we pave the way for developing more sophisticated, efficient, and equitable collaborative systems. These systems will enhance our current visual analytics capabilities and set a precedent for future human-AI interactions across various domains.

\subsection{An Agent-based Framework}

One possible collaborative model is the agent-based framework originally introduced in \cite{monadjemi2023human} and summarized in \autoref{fig:agent_framework}. It draws parallels between human cognitive processes and AI modeling and advocates for a unified language for the visual analytics and \acro{AI} communities. This approach, rooted in the well-established \acro{AI} literature, simplifies complex problems by conceptualizing them as interactions between agents and their environments. Developers and researchers can tailor the specification of this model to their specific contexts and applications.

Applying the agent-based model to visual analytics presents an opportunity to enrich our comprehension and improve the dynamics of human-AI interactions within this domain. In this context, visual analytic agents can be either \textit{human} or \textit{artificial} entities. The model envisions that all agents are capable of both observation and action, contributing toward a collective analytical goal. 

Human agents here are broadly defined and are data scientists, decision-makers, domain experts, or novice users. The prior research on understanding the diverse needs of these groups (e.g.,\ \cite{wong2018towards}) or those that explore how individual differences might influence analytical workflows \cite{liu2020survey,ottley2022adaptive} can inform the model's specifications and considerations.
Additionally, developers can consider studies on how humans perceive data (e.g.,\ \cite{xiong2022seeing,bancilhon2020let}) and the nature of actions undertaken during analytical sessions (e.g.,\ \cite{gotz2009characterizing,brehmer2013multi,gathani2022grammar}).

Artificial agents can consist of modeling algorithms, guidance systems, and automated processes interacting within the environment to assist in collaborative analytical tasks. Prior research in this area has focused on designing artificial agents capable of identifying patterns in data (e.g.\ \cite{kim2019topicsifter,ha2022unified}), learning from user interactions (e.g.,\ \cite{brown2012dis,ottley2019follow}), and assisting users throughout their analytical sessions (e.g.,\ \cite{dabek2016grammar,monadjemi2022guided}). This body of work also highlights the evolving capabilities and contributions of both human and artificial agents in visual analytics, underscoring the potential for synergistic collaboration between these entities in achieving analytical objectives.

This agent-based approach provides a framework for analyzing complex interactions in visual analytics. It also creates opportunities for innovative solutions and advancements in the field. By considering humans and AI systems as agents within a visual analytics environment, we can analyze and improve their interactions, decision-making processes, and information processing in a more effective way.

\section{Discussion and Future Work}
\label{sec:discussion}

The purpose of this paper is to establish the groundwork for the creation of intelligent visual analytics systems that can seamlessly interact with humans. However, given the intricate nature of human behavior and their interaction with AI, there are both obstacles and prospects that need to be addressed to progress in this field. This section will outline the primary challenges that must be overcome to advance this promising area.

\subsection{Understanding Predictable and Unpredictable Human Behaviors}

The advancement of intelligent visual analytic interfaces hinges on their capability to fluidly navigate between handling routine, predictable tasks and engaging with tasks that demand a more intricate and nuanced comprehension of human behavior. The existing body of work, as delineated in \autoref{sec:casestudies}, presents initial examples of predicted actions and tasks. However, this area is still in its early stages of development. The examples in \autoref{sec:casestudies} suggest we can use techniques such as Markov models and active learning algorithms to learn from interactions during data foraging and simple search tasks~\cite{battle2016dynamic,monadjemi2022guided}. Still, examples of predictive algorithms validated with real user data are few and limited~\cite{ha2022unified}. There is still much to do.

To advance this area of research, it is crucial to develop more accurate and reliable predictive algorithms that can be validated with real user data. This will require a deeper understanding of human behavior and the development of more sophisticated data analysis techniques. Nonetheless, intelligent visual analytic interfaces have vast potential benefits and can revolutionize the way we interact with complex data.  Additionally, the community needs to establish protocols for handling situations where the \acro{AI}'s confidence in its predictions is low, as well as expanding the bandwidth of communications between agents. 
Future research is essential for understanding individual variances, how to offer personalized experiences, and how to adjust to users' evolving needs and behaviors. 
Therefore, it is important to continue to invest in this area of research and development to unlock its full potential.
Moving forward in this field requires not just technological advancements, but also a multidisciplinary approach involving psychology, cognitive science, and behavioral studies.

\subsection{Integrating Multi-agent System}

%Explorations into multi-agent systems in visual analytics also hold significant promise. These systems would feature multiple human and artificial agents, each with specialized skills, working in concert with each other. This collaborative approach could lead to more thorough and diverse analytics as various agents contribute their unique expertise to the task. However, this introduces complexities in effectively managing the task allocation and coordination and ensuring that each agent's strengths are utilized effectively. Research in this area must also focus on developing methods for seamless interaction between diverse agents, addressing challenges such as communication protocols, conflict resolution, and decision hierarchy.

With data analysis's growing complexity, more sophisticated tools and techniques are needed to help analysts efficiently and effectively sift through large amounts of data. One promising area of research is multi-agent systems for visual analytics. These systems would feature multiple human and artificial agents, each with specialized skills, working collaboratively to analyze complex data sets.
For example, a human agent could provide the system with domain-specific knowledge, while an artificial agent could perform data processing tasks that are beyond human capability. Another agent could be responsible for data visualization, using its expertise to create interactive and intuitive visualizations, presenting the right data at the right time.

However, managing the task allocation and coordination among different agents presents a significant challenge. It is essential to ensure that each agent's strengths are utilized effectively and that they work together seamlessly. This requires developing methods for seamless interaction between diverse agents.

One of the significant challenges in multi-agent systems is developing communication protocols that enable agents to share information effectively. Conflict resolution is also a critical concern, as different agents may have differing opinions or interpretations of the data. Additionally, it is crucial to establish a decision hierarchy that ensures that each agent's contribution is valued appropriately.

In conclusion, research in multi-agent systems for visual analytics is a promising area that could lead to more effective and efficient data analysis. However, to realize its full potential, it is necessary to address the challenges associated with effectively managing and coordinating multiple agents with diverse skill sets.

\subsection{Addressing Ethical Concerns}

It is crucial for users to trust \acro{AI} algorithms, and transparency in how they function is a key factor in building that trust. This means that algorithms should be designed in a way that is open and clear about how they make decisions and that they can be audited for any biases. One way to make \acro{AI} decision-making more understandable to humans is through the use of Explainable \acro{AI} (\acro{XAI}) techniques.

It is important to make sure that the results produced by \acro{AI} systems are fair and unbiased. This is especially crucial when decisions based on these results can have significant consequences. To achieve fairness, it is necessary to continuously monitor and evaluate the \acro{AI} systems, and identify and address any biases that may arise. Collaborating with experts in ethics, sociology, and relevant fields can provide valuable insights into the societal implications of \acro{AI} decisions, and help create more equitable algorithms.

\section{\uppercase{Conclusions}}
\label{sec:conclusion}

This paper discusses the necessary advancements required to improve intelligent visual analytics systems. We highlight the importance of recognizing the full spectrum of human behavior and examine existing user models that can learn and predict from interaction data. We also suggest expanding the human-machine teaming model and adopting an agent-based model framework that recognizes the potential for collaboration between humans and AI. In addition, we emphasize the need to consider ethical and contextual dimensions while designing such systems, and we discuss other potential future directions. By focusing on these areas, we can create systems that assist and enhance human capabilities in data analysis, embodying a true partnership in the dance of discovery and decision-making.

\section*{\uppercase{Acknowledgements}}

I thank Stefan Jänicke and Helen C. Purchase for inviting me to deliver the keynote speech at \acro{IVAPP} 2024, which is the basis of this manuscript.
I also would like to express my gratitude to Sunwoo Ha for her valuable feedback and to Leilani Battle for allowing the use of her system's imagery.
This material is based upon work supported by the U.S.\ National Science Foundation under grant numbers \acro{IIS}-2142977 and \acro{OAC}-2118201.

\bibliographystyle{apalike}
{\small
\bibliography{example}}

\end{document}